\documentclass[conference]{IEEEtran}
\IEEEoverridecommandlockouts
\usepackage{cite}
\usepackage{amsmath,amssymb,amsfonts}
\usepackage{algorithmic}
\usepackage{graphicx}
\usepackage{textcomp}
\usepackage{xcolor}
\usepackage{enumitem}
\usepackage{booktabs}

\usepackage[linesnumbered,ruled,vlined]{algorithm2e}
\def\BibTeX{{\rm B\kern-.05em{\sc i\kern-.025em b}\kern-.08em
    T\kern-.1667em\lower.7ex\hbox{E}\kern-.125emX}}
\begin{document}

\title{SCALP - Supervised Contrastive Learning for Cardiopulmonary Disease Classification and Localization in Chest X-rays using Patient Metadata\\
% {\footnotesize \textsuperscript{*}Note: Sub-titles are not captured in Xplore and
% should not be used}
% \thanks{Identify applicable funding agency here. If none, delete this.}
}

\author{

\IEEEauthorblockN{Ajay Jaiswal}
\IEEEauthorblockA{UT Austin \\}
\and
\IEEEauthorblockN{Tianhao Li}
\IEEEauthorblockA{UT Austin \\}
\and
\IEEEauthorblockN{Cyprian Zander}
\IEEEauthorblockA{MIS, Germany \\}
\and
\IEEEauthorblockN{Yan Han}
\IEEEauthorblockA{UT Austin \\}
\and
\IEEEauthorblockN{Justin F. Rousseau}
% \IEEEauthorblockA{\textit{Dell Medical School} \\ UT Austin \\}
\IEEEauthorblockA{UT Austin \\}
\and
\IEEEauthorblockN{Yifan Peng}
\IEEEauthorblockA{Weill Cornell Medicine}
\and
\IEEEauthorblockN{Ying Ding}
\IEEEauthorblockA{UT Austin \\}
}

\maketitle

\begin{abstract}
Computer-aided diagnosis plays a salient role in more accessible and accurate cardiopulmonary diseases classification and localization on chest radiography. Millions of people get affected and die due to these diseases without an accurate and timely diagnosis. Recently proposed contrastive learning heavily relies on data augmentation, especially positive data augmentation. However, generating clinically-accurate data augmentations for medical images is extremely difficult because the common data augmentation methods in computer vision, such as sharp, blur, and crop operations, can severely alter the clinical settings of medical images. In this paper, we proposed a novel and simple data augmentation method based on patient metadata and supervised knowledge to create clinically accurate positive and negative augmentations for chest X-rays. We introduce an end-to-end framework, SCALP, which extends the self-supervised contrastive approach to a supervised setting. Specifically, SCALP pulls together chest X-rays from the same patient (positive keys) and pushes apart chest X-rays from different patients (negative keys). In addition, it uses ResNet-50 along with the triplet-attention mechanism to identify cardiopulmonary diseases, and Grad-CAM++ to highlight the abnormal regions. Our extensive experiments demonstrate that SCALP outperforms existing baselines with significant margins in both classification and localization tasks. Specifically, the average classification AUCs improve from 82.8\% (SOTA using DenseNet-121) to 83.9\% (SCALP using ResNet-50), while the localization results improve on average by 3.7\% over different IoU thresholds.  
\end{abstract}


% \begin{figure}[htbp]
% \includegraphics[width=14cm, trim= 29em 10em 5em 8em, clip]{images/overview.pdf}
% \caption{General overview of our chest X-ray image processing framework SCALP for Cardiopulmonary Disease Classification and Localization. SCALP takes the chest X-ray image as the input and estimates prediction probability and renders a heatmap for localization of the disease.}
% \label{overview}
% \vspace{-0.6cm}
% \end{figure}

\section{Introduction}
Chest X-rays (CXR) are one of the most common imaging tools used to examine cardiopulmonary diseases. Currently, CXRs diagnosis primarily relies on professional knowledge and meticulous observations of expert radiologists. Automated systems for medical image classification face several challenges. First, it heavily depends on manually-annotated training data which requires highly specialized radiologists to do manual annotation. Radiologists are already overloaded with their diagnosis duties and their hourly charge is costly. Second, common data augmentation methods in computer vision \cite{chen2020simple, he2020momentum} such as crop, mask, blur, and color jitter can significantly alter medical images and generate inaccurate clinical images. Third, unlike images in the general domain, there is subtle variability across medical images. In addition, a significant amount of the variance is localized in small regions. Thus, there is \textbf{\textit{an unmet need}} for deep learning models to capture the subtle differences across diseases by attending to discriminative features present in these localized regions.

Recently, contrastive learning frameworks which heavily rely on data augmentation techniques \cite{chen2020simple, he2020momentum, chen2020improved} have become promising due to their ability to capture fine-grained discriminative features in the latent space. This paper proposes a novel and simple data augmentation method based on patient metadata and supervised knowledge such as disease labels to create clinically accurate positive samples for chest X-rays. The supervised classification loss helps SCALP create decision boundaries across different diseases while the patient-based contrastive loss helps SCALP learn discriminative features across different patients. Compared to other baselines \cite{rajpurkar2017chexnet, ma2020multilabel}, SCALP uses simpler ResNet-50 architecture and performs significantly better. SCALP uses GradCAM++ \cite{selvaraju2016grad} to generate activation maps that indicate the spatial location of the cardiopulmonary diseases. The highlights of our contribution are:

% The main difference between SCALP and existing approaches is four-fold. \underline{\textit{First}}, SCALP training is single-staged (no finetuning required) using a linear combination of contrastive and supervised classification loss. The patient-based contrastive loss helps SCALP learn discriminative features across different patients simultaneously. The supervised classification loss helps SCALP create decision boundaries across different diseases. \underline{\textit{Second}}, SCALP localization doesn't require usage of any expensive manually annotated bounding box data \cite{li2018thoracic, Liu_2019_ICCV}. In comparison to other baselines \cite{rajpurkar2017chexnet, ma2020multilabel}, SCALP uses simpler ResNet-50 architecture and perform significantly better. \underline{\textit{Third}}, SCALP unconventionally utilizes both patient metadata with supervised knowledge such as disease label, to construct positive and negative keys for the contrastive learning. \underline{\textit{Finally}}, SCALP uses GradCAM++ \cite{selvaraju2016grad} to generate activation maps which indicate the spatial location of the cardiopulmonary diseases. Unlike \cite{li2018thoracic, Liu_2019_ICCV}, we use novel BB-generation algorithm to regular shaped rectangular bounding boxes, instead of irregular regions for localization. 
\begin{itemize}[noitemsep,topsep=0pt,leftmargin=1em]
    
    \item Our augmentation \textbf{technique} for contrastive learning utilizes both patient metadata and supervised disease labels to generate clinically accurate positive and negative keys. \textit{Positive keys} are generated by taking two chest radiographs of the same patient $P$ while \textit{negative keys} are generated using radiographs from patients other than $P$ and having the same disease as $P$.
    \item A novel unified \textbf{framework} to simultaneously improve cardiopulmonary diseases classification and localization. We go beyond the conventional two-staged training (pre-training and fine-tuning) involved in contrastive learning. We demonstrate that single-staged end-to-end supervised contrastive learning can improve existing baselines significantly.
    \item We propose an innovative rectangular \textit{Bounding Box} generation \textbf{algorithm} using pixel-thresholding and dynamic programming. 
\end{itemize}

\section{Related Work}

\subsection{Medical Image Diagnosis}
In the past decade, machine learning and deep learning have played a vital role in analyzing medical data, primarily medical imaging data. Recognition of anomalies and their localization has been a prevalent task for image analysis. Recent surveys \cite{ Anwar_2018, Rezaei2017DeepLF} have illustrated the success of CNNs for classification and localization of several diseases in numerous medical imaging datasets varying from chest X-rays, MRIs, and CT scans. With the availability of large public chest X-rays datasets such as \cite{wang2017chestx, johnson2019mimiccxrjpg}, many researchers have explored the task of thoracic disease classification \cite{seyyed2020chexclusion, rajpurkar2017chexnet, li2018thoracic, Liu_2019_ICCV}. CheXNet \cite{rajpurkar2017chexnet} uses 121-layer CNN trained on ChestXray14\cite{wang2017chestx} for pneumonia detection. \cite{li2018thoracic} and \cite{chaitanya2020contrastive} tried to improve the localization results using manually annotated localization data. % Self-supervised contrastive learning works as a pre-train task by maximizing the agreement of different augmentations of same instance. The origin idea of contrastive learning can date back to 'Becker, Suzanna, and Geoffrey E. Hinton. "Self-organizing neural network that discovers surfaces in random-dot stereograms." Nature 355.6356 (1992): 161-163.', which maximize the agreement of representations of two modules using small size neural network. In these years, a lot of contrastive learning frameworks has been purposed. 'Dosovitskiy, Alexey, et al. "Discriminative unsupervised feature learning with convolutional neural networks." NIPS, 2014.' first treated each instance as a class, now known as instance discrimination, is the core of self-supervised contrastive learning. 'Wu, Zhirong, et al. "Unsupervised feature learning via non-parametric instance discrimination." Proceedings of the IEEE Conference on Computer Vision and Pattern Recognition. 2018.' proposed a memory bank to store the representation of each instance, which improved the performance by enlarge the negative sampling to the whole dataset. 'He, Kaiming, et al. "Momentum contrast for unsupervised visual representation learning." Proceedings of the IEEE/CVF Conference on Computer Vision and Pattern Recognition. 2020.' (MOCO) further boosted the performance by using a queue instead of a memory bank to store the representation of data, which largely sustained consistency among batches. 'Chen, Ting, et al. "A simple framework for contrastive learning of visual representations." International conference on machine learning. PMLR, 2020.' (SIMCLR) purposed a comprehensive self-supervised contrastive learning framework by using multi-augmentation methods, large batch size, and structure improvement (adding a MLP head). A lot of improvement were proposed afterward based on these works. 'SIMCLR v2, MOCO v2, MOCO v3'

\subsection{Contrastive Learning for Chest X-rays}
In the medical domain, prior work has found the performance improvement on applying contrastive learning to the chest X-rays \cite{sowrirajan2021mococxr, sriram2021covid19}. \cite{sowrirajan2021mococxr} presented an adaptation of MoCo \cite{he2020momentum} for chest X-ray dataset by pre-training it on \cite{irvin2019chexpert}.  The closest recent work to our knowledge \cite{vu2021medaug} proposed to use patient metadata did not utilize supervised label information associated with chest X-rays. Recently, \cite{khosla2021supervised} has proposed supervised contrastive learning which extends the self-supervised batch contrastive approach to a fully-supervised setting, allowing us to effectively leverage label information. Inspired by \cite{khosla2021supervised} supervised contrastive loss, we propose to generate data augmentation by exploiting patient data and class labels together.

\section{The Proposed Approach}
Given chest X-rays with the cardiopulmonary disease labels, we aim to design a unified model that simultaneously classifies and localizes cardiopulmonary disease. We formulate both tasks in the same prediction framework and train them using a joint contrastive and supervised binary cross-entropy loss. More specifically, each image in our training data is labeled with an 8-dim vector $y = [y_1,...,y_k, ..., y_K], y_k \in 0,1, K = 8$ for each image, and $y_k$ indicates the presence or absence of 8 cardiopulmonary diseases. Our model produces a probability distribution of over 8 diseases for each image in the test set along with a heatmap with the localization information. The heatmap is passed to the BB-generation algorithm to generate rectangular bounding boxes indicating the presence of pathology.   

\begin{figure}
\includegraphics[width=\columnwidth, trim = 4em 0em 5em 0em, clip]{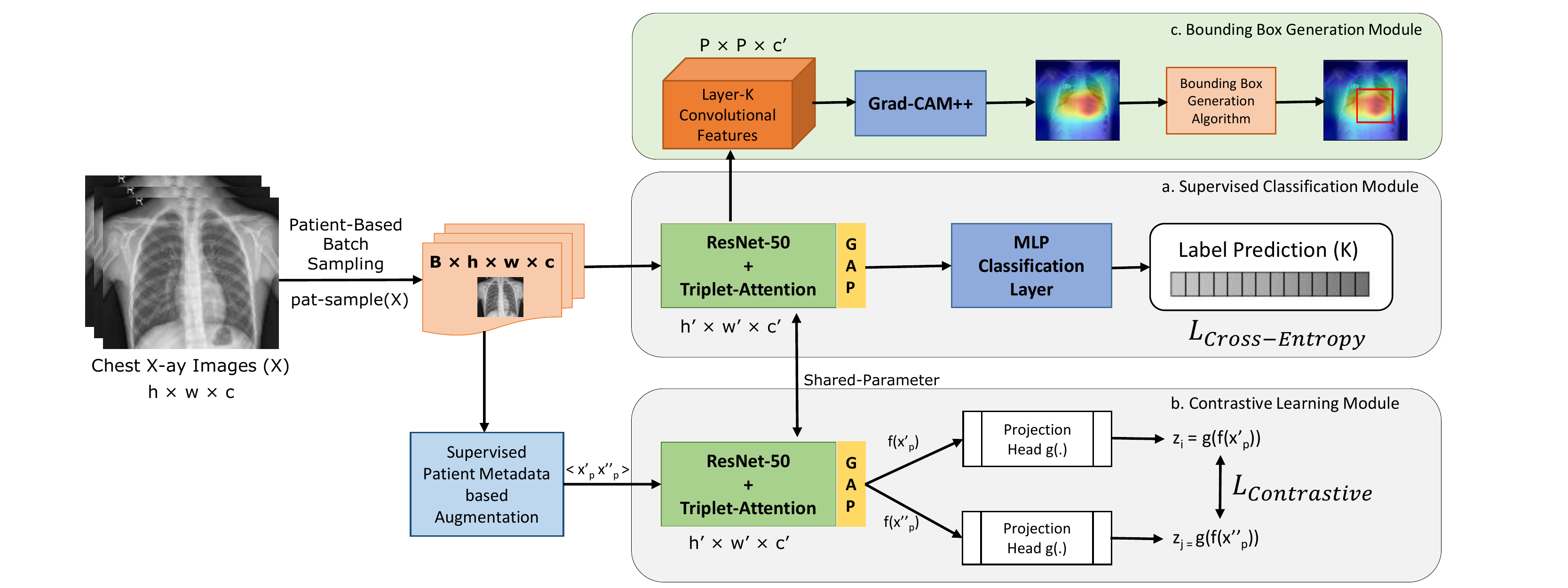}
\caption{Model Overview. The input images are sampled in batches with a constraint that no two images in a batch are from the same patient. Learning is performed using a shared encoder (Resent-50) with triplet attention and joint loss from the supervised classification and contrastive learning module.} 
\label{fig:main_arch}
\end{figure}

\subsection{Image Model}
As shown in Figure \ref{fig:main_arch}, we use the residual neural network (ResNet-50) architecture considering its manageable training with limited GPU resources and popularity in numerous image classification and object detection challenges. Inspired by Triplet attention\cite{misra2020rotate}, we incorporated the  lightweight attention module in our ResNet-50 architecture to use cross-dimension interaction between the spatial dimensions and channel dimension for better localization. Image input with shape $h \times w \times c$ produces a feature tensor with shape $h^{'} \times w^{'} \times c^{'}$, where $h, w,$ and $c$ are the height, width, and the number of channels of the input image, respectively while $h^{'} = h/32, w^{'} = w/32,$ and $c^{'} = 2048$. Our framework is composed of two parallel modules: the supervised classification module and the contrastive learning module. Both modules share the same ResNet-50 encoder, i.e., the same set of parameters for encoding the chest X-ray image inputs.

\subsubsection{Supervised Classification Module:} This module is responsible for learning the high-dimensional decision boundaries across different cardiopulmonary diseases. The encoded input chest X-ray images pass through a global average pooling layer to generate a 2048-dimensional feature vector. The feature vector is fed to a non-linear MLP layer to generate the probability distribution over 8 cardiopulmonary diseases. We calculate $L_{Cross-Entropy}$ by summing loss from each class.
    
\subsubsection{Contrastive Learning Module:} In addition to the inter-class variance of abnormalities in chest X-rays (i.e., feature differences between different diseases which are captured by classification loss), chest X-rays also have a high intra-class variance (i.e., differences in the X-rays of different patients having the same disease). To capture these intra-class variances, we introduce a supervised contrastive learning module to learn discriminative intra-class features. Our Supervised Patient Metadata based Augmentation module (Section \ref{aug}) generates two augmented views $<x'_p\ , x''_p>$ for each image in the batch. After being encoded by a shared encoder $f(.)$, both views are fed to the global average pooling layer to generate feature embedding $f(x'_p)$ and $f(x''_p)$. These feature embeddings are then transformed through the non-linear projection head similar to \cite{chen2020simple} to generate $g(f(x'_p))$ and $g(f(x''_p))$. $L_{Contrastive}$ loss is calculated by maximizing the agreement between $g(f(x'_p))$ and $g(f(x''_p))$\cite{chen2020improved}. 

\subsection{Triplet Attention}
To augment the quality of the localization by exploiting attention from the cross-dimension interaction in feature tensors, we integrate the Triplet Attention \cite{misra2020rotate} into our architecture. Cross-Dimension Interaction involves computing attention weights for each dimension in tensor against every other dimension to capture the spatial and channel attention. In simple terms, spatial attention tells \textit{where the channel to focus on}, while the channel attention tells \textit{what channel to focus on}. With a minimal overload of few learnable parameters, the triplet attention mechanism successfully captures the interaction between the spatial and channel dimension of the input tensor. Following \cite{misra2020rotate}, the input tensor with dimension $H \times W \times C$ in SCALP uses a branching mechanism to capture dependencies between $(C, H), (C, W),$ and $(H, W)$. 

\begin{figure}
\includegraphics[width=\columnwidth, trim = 19em 0em 27em 3em, clip]{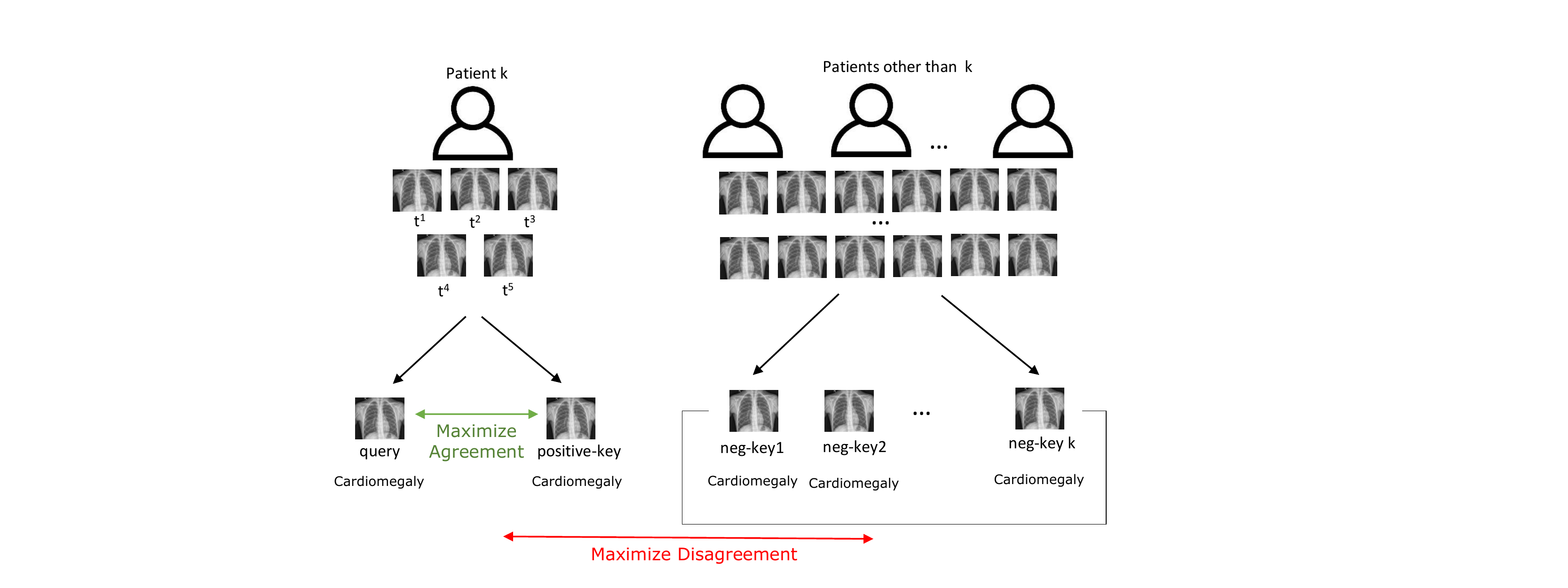}
\caption{Overview of our Supervised Patient-Metadata Based Contrastive Augmentation. Positive and Negative Keys are constructed using Patient Metadata (association of patient id with chest X-rays) and Supervised disease labels of chest X-rays.} 
\label{fig:contrastive}
\end{figure}

\subsection{Supervised Patient-Metadata Based Augmentation}
\label{aug}
Recently proposed Supervised Contrastive Learning \cite{khosla2021supervised} provides an innovative way to leverage label information for generating positive and negative keys effectively. In our case, we have acutely used patient metadata and label information simultaneously to generate positive and negative keys for the SCALP contrastive learning module. The motivation behind introducing contrastive learning in SCALP is to capture intra-class discriminative features between patients diagnosed with the same disease. Figure \ref{fig:contrastive} provides an overview of our novel technique.

\subsubsection{Positive Sampling} Data augmentation in medical imaging is sensitive to augmentation techniques such as random crop, Gaussian blur, and color jitter proposed in prior self-supervised learning \cite{chen2020simple, chen2020improved, he2020momentum}. For medical images, such operations may either change the disease label or are not meaningful for grayscale X-ray images. Since our goal is to incorporate discriminative features between two different patients to SCALP, we randomly select two chest X-ray studies having the same disease label of a patient $P$ using patient metadata. The first study is called \textit{query}, and the second is called a \textit{positive key}. We use the contrastive loss to maximize agreement between them in latent space.

\subsubsection{Negative Sampling} As shown in Figure \ref{fig:contrastive}, we select k negative keys for each patient $P$ in our input batch. \textit{Negative keys} are selected randomly from the pool of chest X-ray studies from patients except $P$, having the same diagnosis as \textit{query}. This helps SCALP to distinguish the subtle differences between patients who are diagnosed with the same disease. Contrastive loss tries to push these \textit{negative keys} away from $query$ during training.    

\begin{table*}[h]
\centering
\begin{tabular}{lccccccccc}
\toprule
Method & Atelectasis & Cardiomegaly & Effusion & Infiltration & Mass & Nodule & Pneumonia & Pneumothorax & Mean\\
\midrule
\textit{Wang et. al.}\cite{wang2017chestx} &  0.72 & 0.81 & 0.78 & 0.61 & 0.71 & 0.67 & 0.63 & 0.81 & 0.718\\
\textit{Wang et. al.}\cite{wang2018tienet} &  0.73  & 0.84  & 0.79    & 0.67  & 0.73  &  0.69 &  0.72  &  0.85 & 0.753\\
\textit{Yao et. al.}\cite{yao2017learning} &  0.77 & 0.90 & 0.86 & 0.70 & 0.79 & 0.72 & 0.71 & 0.84 & 0.786\\
\textit{Raj. et. al.}\cite{rajpurkar2017chexnet} & 0.82 & 0.91 & 0.88 & 0.72 & 0.86 & 0.78 & 0.76 & 0.89 & 0.828\\
\textit{Kum. et. al.}\cite{kumar2018boosted} & 0.76 & 0.91 & 0.86 & 0.69 & 0.75 & 0.67 & 0.72 & 0.86 & 0.778\\ 
\textit{Liu et. al.}\cite{Liu_2019_ICCV} & 0.79 & 0.87 & \textcolor{red}{0.88} & 0.69 & 0.81 & 0.73 & 0.75 & 0.89 & 0.801\\
\textit{Seyed et. al.}\cite{seyyed2020chexclusion} & \textcolor{red}{0.81} & 0.92 & 0.87 & 0.72 & 0.83 & 0.78 & 0.76 & 0.88 & 0.821\\ 
\midrule
Our model             & 0.79  & \textcolor{red}{0.92}  & 0.79  & \textcolor{red}{0.89}  & \textcolor{red}{0.88}  & \textcolor{red}{0.87}  & \textcolor{red}{0.77}  & \textcolor{red}{0.81}  & \textcolor{red}{0.839}\\
\hspace{2em}(std) & $\pm 0.01$ & $\pm 0.00$ & $\pm 0.01$ & $\pm 0.01$ & $\pm 0.02$ & $\pm 0.00$ & $\pm 0.01$ & $\pm 0.02$\\
\bottomrule
\end{tabular}
\vspace{0.3cm}
\caption{\label{table1}
Comparison with the baseline models for AUC of each class and average AUC. For each column, red values denote the best results. Note that the best baseline with mean AUC 0.828 uses the DenseNet-121 architecture, while our model is trained using a comparatively simple and light-weight ResNet-50 architecture.
}
\label{tab:classificationresult}
\vspace{-0.5cm}
\end{table*}
\subsection{Loss Function}
SCALP is trained using the linear combination of supervised classification and contrastive loss. For the supervised classification, we use binary cross-entropy loss. For contrastive learning, we use the extended version of NT-Xent loss\cite{chen2020simple}.

\subsubsection{Supervised classification Loss} Our disease classification is a multi-class classification problem where we have 8 disease types. Multiple diseases can often be identified in one chest X-ray image and diseases are not mutually exclusive. We, therefore, define 8 binary classifiers for each class/disease type. Since all images in our dataset have 8 labels, the loss function for class k can be expressed as minimizing the binary cross-entry as:
\begin{equation}
    L_k = - y_k \cdot log p(k|I) - (1-y_k)\cdot log(1 - p(k|I))
\end{equation}
where $y_k$ is the ground-truth label of the $k$-th class, and $I$ is the input image. To enable end-to-end training across all the classes, we sum up the class-wise loss to calculate total supervised loss as:
\begin{equation}
    L_{Cross-Entropy} = \sum_{n=1}^{batch-size}\sum_{k=1}^{8} L^n_k
\end{equation}

\subsubsection{Contrastive Loss} Our contrastive loss extends the normalized temperature scaled cross-entropy loss (NT-Xent). Using our patient-based sampling, we randomly select a batch of $N$ chest X-ray images belonging to $N$ patients. We derive the contrastive loss on the pairs of augmented examples from the batch. Let $I'_{PD}$ be an image in batch belonging to patient $P$ with disease $D$, $sim(x, y)$ denotes similarity between $x$ and $y$, and $f(.)$ denotes ResNet-50 encoder, $g(.)$ denotes the projection head. The loss function $l(I'_{PD})$ for a positive pair of example $<I'_{PD}, I''_{Pd}>$ is defined as: 

\begin{equation}
    l(I'_{PD}) = -log \frac{e^{(sim(g(f(I'_{PD})),\ g(f(I''_{PD})))/\tau)}}{\sum_k \mathbb{1}_{[p \neq P,\ d = D]} e^{(sim(g(f(I'_{PD})),\ \\g(f(I''_{pd})))/\tau)}}
\end{equation}
where $\mathbb{1}_{[p \neq P,\ d = D]} \in$ \{0, 1\} is an indicator function evaluating to 1 iff $p \neq P$ and $d = D$, $\tau$ is the temperature parameter. The final contrastive loss is calculated as the sum over all instances in batch:
\begin{equation}
    L_{Contrastive} = \sum_{k=1}^{batch-size} l(I^k_{PD})
\end{equation}

Eventually, we treat SCALP learning as the optimization of both contrastive and supervised cross-entropy loss together. Total loss for SCALP is defined as:
\begin{equation}
    L_{Total} = \lambda \times L_{Cross-Entropy} + (1 - \lambda) \times L_{Contrastive}
\end{equation}

\subsection{Bounding Box Generation Algorithm}
\label{BB-gen}
We propose an innovative and time-efficient approach $\mathcal{O}(n^2)$ to generate regular-shaped rectangular \textit{bounding boxes} on chest X-rays indicating the approximate spatial location of the predicted cardiopulmonary disease. As shown in Figure \ref{fig:main_arch}, we feed the k-th layer of our image encoder (ResNet-50) to Gradient-weighted Class Activation Mapping (GradCAM++)\cite{selvaraju2016grad} to extract the attention maps/heatmaps. Due to the simplicity of intensity distributions in these heatmaps, we first scale heatmaps to the range [0, 255] and apply an ad-hoc threshold to convert these heatmaps into a binary matrix. Pixel values are converted to 1 if its intensity is greater than the threshold and 0 otherwise. Many previous works \cite{wang2017chestx, yanradiomics} use only intensity threshold to generate bounding boxes which lead to many false positives. We use dynamic programming to generate a set of k candidate rectangles for bounding boxes and eliminate false positives by selecting the candidate which has the highest average intensity per pixel.

\section{Experiments}
\begin{table*}
\centering
\begin{tabular}{lcccccccccc}
\toprule
\textbf{T(IoU)} & Model & \textbf{Atelectasis} & \textbf{Cardiomegaly} & \textbf{Effusion} & \textbf{Infiltration} & \textbf{Mass} & \textbf{Nodule} & \textbf{Pneumonia} & \textbf{Pneumothorax} & \textbf{Mean}\\
\midrule
%&&& \multicolumn{3}{c}{\textbf{T(IoU) = 0.1}} &&&\\
%\hline
0.1 & Baseline \cite{wang2017chestx} & 0.69 & 0.94 & 0.66 & 0.71 & 0.40 & 0.14 & 0.63 & 0.38 & 0.569\\
 & Our Model & 0.62 & 0.97 & 0.64 & 0.81 & 0.51 & 0.12 & 0.8 & 0.29 & \textcolor{red}{0.595} \\
%\hline
%&&& \multicolumn{3}{c}{\textbf{T(IoU) = 0.2}} &&&\\
\midrule
0.2 & Baseline \cite{wang2017chestx} & 0.47 &0.68 &0.45& 0.48& 0.26 &0.05 &0.35 &0.23& 0.371\\
 & Our Model & 0.42 & 0.92 & 0.42 & 0.6 & 0.25 & 0.04 & 0.56 & 0.18 & \textcolor{red}{0.434} \\
%\hline
%&&& \multicolumn{3}{c}{\textbf{T(IoU) = 0.3}} &&&\\
\midrule
0.3 & Baseline \cite{wang2017chestx} & 0.24 &0.46 &0.30& 0.28& 0.15& 0.04& 0.17 &0.13 &0.221\\
 & Our Model & 0.29 & 0.78 & 0.23 & 0.37 & 0.13 & 0.01 & 0.4 & 0.05& \textcolor{red}{0.283} \\
%\hline
%&&& \multicolumn{3}{c}{\textbf{T(IoU) = 0.4}} &&&\\
\midrule
0.4 & Baseline \cite{wang2017chestx} & 0.09& 0.28& 0.20& 0.12& 0.07& 0.01& 0.08& 0.07& 0.115\\
 & Our Model & 0.18 & 0.55 & 0.12 & 0.19 & 0.09 & 0.01 & 0.25 & 0.02 & \textcolor{red}{0.176}\\
%\hline
%&&& \multicolumn{3}{c}{\textbf{T(IoU) = 0.5}} &&&\\
\midrule
0.5 & Baseline \cite{wang2017chestx} & 0.05 &0.18& 0.11& 0.07& 0.01& 0.01& 0.03& 0.03& 0.061\\
 & Our Model & 0.07 & 0.33 & 0.04 & 0.10 & 0.04 & 0.0 & 0.14 & 0.10 & \textcolor{red}{0.102} \\
%\hline
%&&& \multicolumn{3}{c}{\textbf{T(IoU) = 0.6}} &&&\\
\midrule
0.6 & Baseline \cite{wang2017chestx} & 0.02 &0.08 &0.05 &0.02 &0.00 &0.01 &0.02 &0.03 &0.029\\
 & Our Model & 0.02 & 0.14 & 0.02 & 0.04 & 0.03 & 0.00 & 0.07 & 0.00 & \textcolor{red}{0.040} \\
%\hline
%&&& \multicolumn{3}{c}{\textbf{T(IoU) = 0.7}} &&&\\
\midrule
0.7 & Baseline \cite{wang2017chestx} & 0.01 &0.03 &0.02 &0.00& 0.00 &0.00 &0.01& 0.02& 0.011\\
 & Our Model & 0.01 & 0.04 & 0.01 & 0.03 & 0.01 & 0.00 & 0.02 & 0.00 & \textcolor{red}{0.015} \\
\bottomrule
\end{tabular}
\vspace{0.3cm}
\caption{\label{table1}
Disease localization under varying IoU on the NIH Chest X-ray dataset. Note that since our model doesn’t use any ground truth bounding box information, to fairly evaluate the performance of our model, we only consider the previous methods’ results with the same settings as SCALP.
}
\label{tab:localizationresult}
\vspace{-0.1cm}
\end{table*}

\subsection{Dataset and Preprocessing}
NIH Chest X-ray dataset \cite{wang2017chestx} consists of 112,120 chest X-rays collected from 30,805 patients, and each image is labeled with 8 cardiopulmonary disease labels. The NIH dataset also includes high-quality bounding box annotations for 880 images by radiologists. We separate these 880 images from our entire dataset, and they are used only to evaluate disease localization. Our method does not require any training data related to bounding boxes which is a significant difference compared to other existing baseline methods \cite{Liu_2019_ICCV, li2018thoracic} which use some percentage of these images for training. We follow the same protocol as \cite{li2018thoracic}, to shuffle our dataset (excluding images with BB annotations) into three subsets: 70\% for training, 10\% for validation, and 20\% for testing. In order to prevent data leakage across patients, we make sure that there is no patient overlap between our subsets.

\begin{algorithm}
\DontPrintSemicolon
  \textbf{Input:} k-th layer attention map/heatmap from ResNet-50\\
  \textbf{Output:} coordinates (x1, y1, x2, y2) of the bounding box \\
  \textbf Scale heatmap intensities to [0, 255] and create a mask matrix with the same dimension as the heatmap.\\
   \If {pixel $>$ 180}
    {
       mask[pixel] = 1\\
    }
    \Else
    {
        mask[pixel] = 0
    }
    \textbf Using dynamic programming \cite{dp}, generate k maximum area rectangles as candidate BB.\\
    \textbf Expand candidate rectangles uniformly across  the edge till newly added \textit{ratio (0s count, 1s count)} $>$ 1 \\
    \textbf Select the rectangle with the maximum average pixel intensity mapped in the heatmap and return its coordinates.
\caption{Bounding Box Generation Algorithm}

\end{algorithm}

\subsection{Implementation Details} 
We used the ResNet-50 model with triplet attention and initialized it with pre-trained weights provided by \cite{misra2020rotate}. Our MLP layer is a two-layered fully-connected network with RELU non-linearity, while the projection head is defined similarly to \cite{chen2020simple}. Both our projection head and MLP layer are randomly initialized. We have used 0.01 learning rate and weight decay of $10^{-6}$ and $10^{-4}$ for contrastive and classification loss, respectively. SCALP uses an SGD optimizer and learning rate scheduler with step size and gamma value of 10 and 0.1, respectively.
\begin{table}
\centering
\begin{tabular}{lcc}
\toprule
\textbf{} & \textbf{SCALP w/o Contrastive}& \textbf{SCALP}\\
\midrule
Atelectasis & 0.751 & 0.79\\
Cardiomegaly & 0.850 & 0.92\\
Effusion & 0.833 & 0.79\\
Infiltration & 0.670 & 0.89\\
Mass & 0.694 & 0.88\\
Nodule & 0.640 & 0.87\\
Pneumonia & 0.700 & 0.77\\
Pneumothorax & 0.792 & 0.81\\
\midrule
\textbf{Mean} & 0.7413 & 0.839\\
\bottomrule
\end{tabular}
\vspace{0.3cm}
\caption{AUC comparison of SCALP with and without contrastive learning module. 
%A significant drop in performance illustrates the benefits of our contrastive learning module.
}
\label{tab:params}
\vspace{-0.2cm}
\end{table}

\begin{table}
\centering
\begin{tabular}{c|cccccc}
\toprule
\textbf{$\lambda$} & 0.99 & 0.90 & 0.85 & 0.80 & 0.75 & 0.70 \\
\midrule
\textbf{AUC} & 0.766 & 0.794 & 0.822 & \textbf{0.839} & 0.818 & 0.785 \\
\bottomrule
\end{tabular}
\vspace{0.3cm}
\caption{AUC comparison of SCALP for varying $\lambda$ in Equation 7. A higher value of $\lambda$ implies lower weight to the contrastive loss. SCALP achieves the best performance when 80\% weight is given to classification loss and 20\% weight is given to contrastive loss.}
\label{tab:interpolation}
\vspace{-2.6em}
\end{table}

\subsection{Disease Identification}
SCALP classification is a multi-label classification problem. It assigns one or more labels among 8 cardiopulmonary diseases. We conduct a 3-fold cross-validation (Table \ref{tab:classificationresult}). We compare SCALP with reference models, which have published state-of-the-art performance of disease classification on the NIH dataset. We have used Area under the Receiver Operating Characteristics (AUC) to estimate the performance of our model in Table \ref{tab:classificationresult}. Our results also present the 3-fold cross-validation to show the robustness of our model. Compared to other baselines, SCALP achieves a mean AUROC score of 0.839 using ResNet-50 across the 8 different classes, which is 0.011 higher than the SOTA (uses DenseNet-121) on disease classification.

To understand the importance of contrastive module for disease classification, we trained SCALP with and without the contrastive loss and evaluated performance on the test set of NIH data. Table \ref{tab:params} presents the significant drop of $> 9$ \% AUC when we exclude the contrastive module from the SCALP pipeline. This demonstrates the importance of our innovative association of contrastive modules with the classification pipeline. Our experiments in Table \ref{tab:interpolation} prove our hypothesis that both contrastive and cross-entropy loss is important. In a calculated ratio, they help SCALP learn both disease-level and patient-level discriminative visual features.

\subsection{Disease Localization}
The NIH dataset has 880 images labeled by radiologists with the bounding box information. We have used this dataset to evaluate the performance of SCALP for disease localization. Many prior works \cite{li2018thoracic, Liu_2019_ICCV} have used a fraction of ground truth (GT) bounding boxes for training and evaluated their system on the remaining. To ensure a robust evaluation, we do not use any GT for training, and Table \ref{tab:localizationresult} presents our evaluation results on all 880 images. For localization, we evaluated our detected regular rectangular regions against the annotated ground truth (GT) bounding boxes, using intersection over union ratio (IoU). The localization is defined as correct only if IoU $>$ T(IoU). We evaluate SCALP for different thresholds ranging from \{0.1, 0.2, 0.3, 0.4, 0.5, 0.6, 0.7\} as shown in Table \ref{tab:localizationresult}. A higher  IoU threshold is preferred for disease localization because clinical usage requires high accuracy. Note that SCALP mean performance for 8 diseases is significantly better than the baseline under all IoU thresholds. When the IoU is set to 0.1, SCALP outperforms the baseline in terms of Cardiomegaly,  Infiltration, Mass, and Pneumonia. 

Note that our innovative bounding box generation algorithm successfully eliminates dispersed attention, and identifies regions where maximum attention is concentrated. For example, in "Effusion", the generated heatmap has dispersed attention on both sides of the lungs. However, attention intensity is concentrated in the left side of the lung. Our algorithm is able to generate a bounding box on the left side of the lung and have a high overlap with the ground-truth. Similarly, for "Infiltration" and "Nodule", many undesirable patches of attention have been eliminated which is helpful in improving the IoU evaluation of SCALP. The attention maps generated by SCALP are sharp and focused compared to our reference baseline \cite{wang2017chestx}. Overall, our results show that the predicted disease localizations have significant alignment with the ground truth and can serve as interpretable cues for the disease classification.

\section{Conclusion}
In this work, we propose a simple and effective end-to-end framework SCALP using supervised contrastive learning to identify cardiopulmonary diseases in chest X-ray. We go beyond two-stage training (pre-training and fine-tuning), and demonstrate that an end-to-end supervised contrastive training using two images from the same patient as a positive pair, can significantly outperform SOTA on disease classification. SCALP can jointly model disease identification and localization using the linear combination of contrastive and classification loss. We also propose a time-efficient Bounding Box generation algorithm that  generates bounding boxes from the attention map of SCALP. Our extensive qualitative and quantitative results demonstrate the effectiveness of SCALP and its state-of-the-art performance.

\section{Acknowledgment}

This work is supported by Amazon Machine Learning Research Award 2020. It also was supported by the National Library of Medicine under Award No. 4R00LM013001.

% This work is supported by Amazon Machine Learning Research Award 2020.
% \subsubsection{Batch Size}
% \subsubsection{Interpolation parameter $\alpha$}
% \subsubsection{Loss Function} - focal loss, BCE loss, Margin Loss
% \subsubsection{Heatmap across k different channels for Grad Cam visualization}

\begin{figure}
\centering
\includegraphics[width=9cm,trim=0em 24em 4em 4em ]{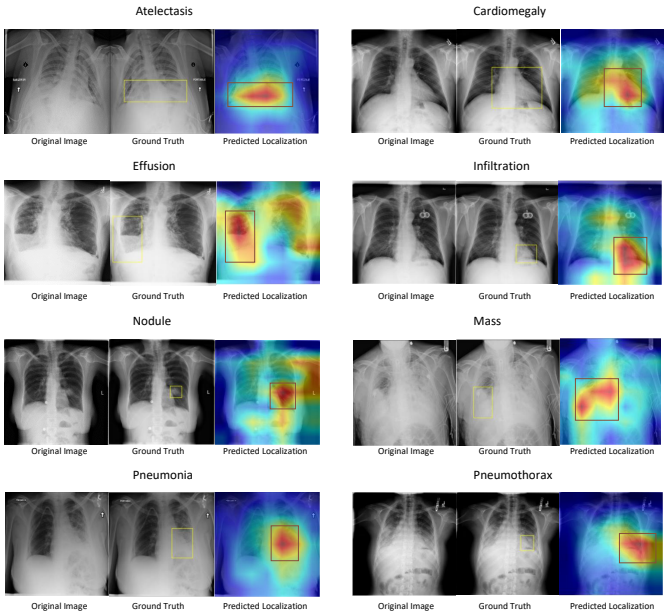}
\caption{Examples of visualization of localization on the test images. We plot the results of diseases near the thoracic. The attention maps are generated from the fourth layer of SCALP's encoder and overlapped with its corresponding original radiology image. The ground-truth and the predicted bounding boxes are shown in yellow and red color respectively.}
\label{fig:heatmap}
\end{figure}

%\vspace{-0.5cm}

\begin{figure}
\includegraphics[width=8cm, trim = 0em 2em 0em 0em, clip]{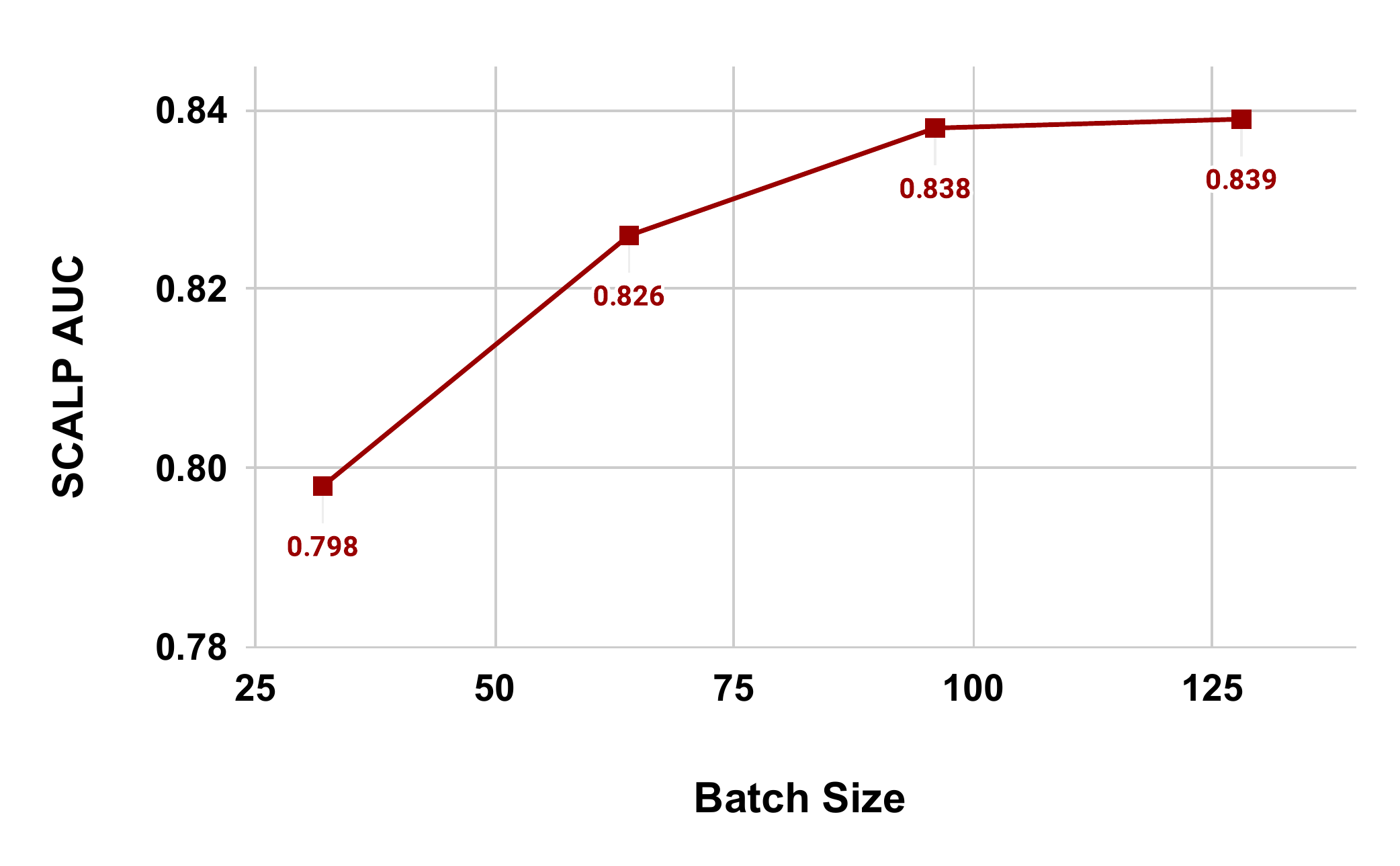}
\caption{Effect of Batch Size on SCALP performance for Disease Classification. Prior works\cite{chen2020simple, he2020momentum} have verified that contrastive learning benefits from a larger batch size. SCALP shows a similar trend with increasing batch size. } 
\label{fig:batch_plot}
\end{figure}
%\vspace{-0.4cm}
% \begin{thebibliography}{00}
% \bibitem{b1} G. Eason, B. Noble, and I. N. Sneddon, ``On certain integrals of Lipschitz-Hankel type involving products of Bessel functions,'' Phil. Trans. Roy. Soc. London, vol. A247, pp. 529--551, April 1955.
% \bibitem{b2} J. Clerk Maxwell, A Treatise on Electricity and Magnetism, 3rd ed., vol. 2. Oxford: Clarendon, 1892, pp.68--73.
% \bibitem{b3} I. S. Jacobs and C. P. Bean, ``Fine particles, thin films and exchange anisotropy,'' in Magnetism, vol. III, G. T. Rado and H. Suhl, Eds. New York: Academic, 1963, pp. 271--350.
% \bibitem{b4} K. Elissa, ``Title of paper if known,'' unpublished.
% \bibitem{b5} R. Nicole, ``Title of paper with only first word capitalized,'' J. Name Stand. Abbrev., in press.
% \bibitem{b6} Y. Yorozu, M. Hirano, K. Oka, and Y. Tagawa, ``Electron spectroscopy studies on magneto-optical media and plastic substrate interface,'' IEEE Transl. J. Magn. Japan, vol. 2, pp. 740--741, August 1987 [Digests 9th Annual Conf. Magnetics Japan, p. 301, 1982].
% \bibitem{b7} M. Young, The Technical Writer's Handbook. Mill Valley, CA: University Science, 1989.
% \end{thebibliography}
% \vspace{12pt}
% \color{red}
% IEEE conference templates contain guidance text for composing and formatting conference papers. Please ensure that all template text is removed from your conference paper prior to submission to the conference. Failure to remove the template text from your paper may result in your paper not being published.

\bibliographystyle{./IEEEtran}
\bibliography{./conference_101719}

\end{document}